\documentclass[fleqn,twoside]{article}
\usepackage{espcrc2}

\usepackage{epsfig}
\usepackage{psfig}

\newcommand{\bmat}{\left(\begin{array}}
\newcommand{\emat}{\end{array}\right)}
\newcommand{\be}{\begin{equation}}
\newcommand{\ee}{\end{equation}}
\newcommand{\bea}{\begin{eqnarray}}
\newcommand{\eea}{\end{eqnarray}}
\def\ie{{\it i.e.}}
\def\lsim{\raise0.3ex\hbox{$\;<$\kern-0.75em\raise-1.1ex\hbox{$\sim\;$}}}
\def\gsim{\raise0.3ex\hbox{$\;>$\kern-0.75em\raise-1.1ex\hbox{$\sim\;$}}}
\def\Frac#1#2{\frac{\displaystyle{#1}}{\displaystyle{#2}}}

\title{ 
{\em \small Expanded version of the contribution in 
 NEUTRINO2002, May 25-30,2002, Munich, Germany}\\[1cm]
\bf Neutralino relic density from direct non-equilibrium
 production and intermediate  scales}

\author{E. Torrente-Lujan
\thanks{The work presented here was made in 
collaboration with S. Khalil and C. Munoz, see. Ref.\protect\cite{torrenterelic}} 
\address{Dept.  F\'{\i}sica
Te\'orica C-XI, Univ. Aut\'onoma de Madrid, 
Cantoblanco, 28049 Madrid, Spain.}
}

\begin{document}

\begin{abstract}

We review the calculation of the 
 LSP relic density in alternative cosmological 
scenarios where contributions from direct decay production 
are important. We study  supersymmetric models with
intermediate unification scale. We find concrete scenarios where 
the reheating
temperature is of order one GeV ($M_I\sim 10^{12}$) and below.
If the case that this reheating temperature is associated to 
the decay of oscillating moduli fields  
appearing in string theories, 
we show that the LSP relic  density considerably 
increases with respect to 
the standard radiation-dominated case by 
the effect of the direct non-thermal 
production by the modulus field. The LSP can become a  
good dark matter
candidate ($0.01-0.1 \gsim \Omega h^2 \lsim 0.3-1$) 
for $M_I\sim 10^{12}-10^{14}$ Gev and $m_\phi\sim 1-10$ TeV.
\end{abstract}
\maketitle

\section{Introduction}

%%\bibitem{relicfebrero} khalil carlos torrente

In  a series of works \cite{torrenterelic,interm1,interm2}, various
implications for an  intermediate unification scale in
supersymmetric models have been considered. In particular, it was
pointed out that the lightest neutralinos (LSP)--nucleon cross
sections are very sensitive to the variation of this scale. For
instance, for $M_I \sim 10^{12}-10^{14}$ GeV the cross sections are
enhanced and a large region of the parameter space of the minimal
supersymmetric standard model (MSSM) is compatible with the
sensitivity of the current and near future 
direct detection 
dark matter experiments (DAMA, CDMS, EDELWEISS). 
In fact, such
results can also be obtained within the GUT unification 
scenario,
\ie, $M_I \sim 10^{16}$ GeV, but it requires 
tuning the available parameter space:
either large $\tan
\beta$ ($\tan \beta \ge 25$) or a 
specific choice for non--universal
soft SUSY breaking terms so that the value of $\mu$ term is
reduced and hence a larger Higgsino components of the LSP is
obtained, which is essential for enhancing the LSP--nucleon cross
sections \cite{interm1}.

One problem is than in GUT unification scenarios with 
 standard Big Bang (BB) cosmological expansion
where the LSP is produced in a thermal environment dominated 
by radiation,
 such large cross section
leads to a very small LSP relic density 
(outside the cosmological and astrophysical
interesting region $0.1 \le \Omega h^2 \le 0.3$).
In this scheme there is a generic anti-correlation  between
the LSP relic density and its quark or 
nucleon cross sections:
 one usually obtains 
\be \Omega_{\chi} h^2
\simeq \Frac{C}{\langle \sigma^{\mathrm{ann}}_{\chi} v \rangle }~,
\ee 
where $\sigma^{\mathrm{ann}}$ is the annihilation cross 
section for
 neutralino pair, 
$v$ is the pair relative velocity, 
and $\langle .. \rangle$ denotes thermal averaging.
%%%The constant $C$ involves factors of Newton's constant, the
%%%%%%temperature of cosmic background radiation, etc. 
Scattering ($\sigma_{\chi-q}$) and annihilation 
$\sigma^{\mathrm{ann}}_{\chi}$ 
LSP cross sections can be related by a cross symmetry. 
Therefore, a large
scattering cross section $\sigma_{\chi-q}$ leads 
generically to a large
annihilation cross section and a small relic density 
as Eq. 1 suggest.
Indeed, in these models, it was observed that LSP--proton cross
sections of order $10^{-6}$ Pb correspond to too 
low relic densities
($<\sim 0.005-0.01$ ).

We want to address here the relic density question: in the 
context of diverse non-standard 
cosmological scenarios, what is  the effect of changing the
unification scale from GUT scale ($\sim 10^{16}$ GeV) down to
intermediate scale ($\sim 10^{12}$ GeV) and whether
 we still have any  anti-correlation between $\Omega_{\chi} h^2$
and $\sigma^{\mathrm{ann}}_{\chi}$.

\section{The standard BB scenario}
The LSP relic density is given by 
$$\Omega_{\chi} h^2 =\rho_{\chi}/\rho_{crit},$$ 
where $\rho_{\chi}$ is the LSP mass density and 
$\rho_{crit}= 3 H_0^2/8 \pi G \simeq 1.9 \times
10^{-29} h^2 \ g/cm^3$. The evolution of the number density
$n_{\chi} = \rho_{\chi}/m_{\chi}$ can be described to a 
precision sufficient for 
our purposes by the Boltzmann
equation 
\be \frac{d n_{\chi}}{d t} + 3 H n_{\chi} = - \langle
\sigma^{ann}_{\chi} v \rangle \left[ (n_{\chi})^2
-(n_{\chi}^{eq})^2 \right], 
\label{boltzmann} 
\ee 
where $H$ is the Hubble expansion rate. 
In the standard calculation, 
%%%%where very heavy reheating temperature is considered, 
it is assumed  that in the early universe $(T \gg m_{\chi}$) 
the LSP density was initially equal to its  
equilibrium value $n_{\chi} =n_{\chi}^{eq}$.
As the universe expands the LSP density traces its 
equilibrium value. When the particle becomes non relativistic 
( $T < m_{\chi}$) 
the number density of $\chi$'s drops exponentionally.  The
annihilation rate 
$\Gamma_{\chi} \equiv \langle\sigma^{ann}_{\chi} v \rangle n_{\chi}$ 
becomes smaller than the
expansion rate, $\Gamma_{\chi} \lsim H$. 
At this point, the LSP  decouples and
fall out from being in equilibrium with the 
radiation, their number remains constant after that and its density is 
only diluted by the universe expansion.

As long as the LSP is  non-relativistic, we can expand the 
averaged cross section 
$\langle \sigma^{ann}_{\chi} v \rangle$ 
as follows 
\be 
\langle \sigma^{ann}_{\chi} v \rangle = \alpha_s +\alpha_p 
\langle v^2\rangle, 
\label{sigma} 
\ee 
where $\alpha_s$ is the s-wave contribution at zero relative velocity 
and $\alpha_p$ contains contribution from both s- and p- waves. 
The relic density solution to the Boltzamn equation 
is  accurately given by  the 
approximate formula (c.f. Eq.1)
\be 
\Omega_{\chi} h^2 \simeq  \Frac{8.8 \times 10^{-11} GeV^{-2}}
{\surd g_*  (\alpha_s/x_F + 3 \alpha_p/x_F^2)}, 
\label{omega1}
 \label{thermal2} 
\ee 
where the freezing-out temperature $x_F \equiv  m_{\chi}/T_F$ 
can be estimated as 
\be 
x_F \sim \ln \Frac{c  
(\alpha_s +6\alpha_p/x_F) }{\sqrt{g_* x_F}}. 
\ee 
In these formulas 
%$M_P = 1.22\times 10^{19}$ GeV is the Planck mass, 
$g_*$ is the effective number of
relativistic degrees of freedom at $T_F$ and $c$ is constant of
order $10^{-1}$ \cite{report}.

\section{Alternative scenarios}
In alternative cosmological scenarios,
it is often assumed the existence of one (o more) epoch before the radiation-dominated era 
of the universe where the energy is dominated by coherent oscillating fields, the so called reheating era.
A reheating process is associated by example
 with the final stage of Inflation. It can be also, 
independently or not, associated 
with the oscillations of moduli or Polony fields appearing 
in string theory after 
flat directions acquire a mass from SUSY breaking. 

The details and physical origin of these reheating are 
largely unimportant. Its effects can be characterized
 by a reheating temperature $T_{RH}$ which is   conventionally defined (see the definition below)
 as the temperature
  at which the oscillating
field energy  ceases to dominate the cosmological evolution and 
starts the radiation dominated epoch.

The standard BB scenario presented in the previous section
can be seen as an special 
case when $T_{RH}$ is too much higher than $T_F$ 
(of order $10^{9-10}$ GeV ). In this case, the reheating epoch
has no relevance in the final output of the relic density.

A problem could appear if the decay of the oscillating field 
occurs after or during the primordial BBN nucleosynthesis.
 This is drastically solved imposing from the onset that the 
reheating temperature is larger than $\sim 1$ MeV.

Finally, in interesting intermediate cases,
a relatively low value of $T_{RH}$ such that $T_F>T_{RH}$ 
can have qualitative and quantitative 
implications on the predictions of the
relic abundance of the LSP as discussed in Ref. \cite{kolb}. 
The main reason for this is that now  the decoupling of 
the LSP particle occurs in  
an expanding universe dominated by matter instead than 
radiation and that has effects.
We will see below that theories with intermediate 
unification scale
predict low reheating temperature ($\lsim GeV$ and below)
 and they can be considered as interesting scenarios.

%%%\section{others}

It is useful to use the following  expression which 
can be interpreted as a quantitative definition 
of the temperature $T_{RH}$ and which   
 connects it to the decay width of the
 oscillating field:
\be 
\Gamma_\phi = \left(\frac{4 \pi^3 g_*(T_{RH})}{45}\right)^{-1/2} 
\frac{T_{RH}^2}{M_P}.
\label{trh}
\ee 
Given any concrete theory, the reheating temperature can be related to 
the their physical parameters  inverting Eq.\ref{trh} for $T_{RH}$ as 
we can see on the following.

Let us consider first an inflation scenario. 
The inflaton $(\phi)$ decay width which $\Gamma_\phi$ 
can be written, in some
some supersymmetric
hybrid inflation which has been 
considered \cite{george}, as 
$$\Gamma_{\phi}= \frac{1}{8 \pi }\left ( \frac{M_{f}}{\langle \phi \rangle}\right )^2  m_{\phi},$$
 where $m_{\phi}$ is the inflaton mass.
Here $M_{f}$ is the mass of the particle $f$ that the inflaton 
decay to (i.e. a  right handed neutrino or a sneutrino). 
$M_{f}$ should be less than the inflaton mass to allow 
for the decay 
$\phi \to f f$. 
Finally $\langle \phi \rangle $ is the vev of the inflaton field 
which is of the order of the unification scale. 

In Ref. \cite{linde} it
has been shown that a intermediate unification scale of order
$M_I\sim 10^{11-12}$ GeV is favored by  inflation. 
In this case, 
recalling that the inflaton mass is constrained by 
$$m_{\phi} < M_I^2 / M_P,$$
 one obtains
($\langle \phi \rangle\sim 1.7\times 10^{16}$ GeV \cite{george})
 an inflaton mass of the order  $m_{\phi}\sim 10^2$ GeV.  For the 
'standard GUT scenario' with $M_I \sim 10^{16}$ 
one obtains however  $m_{\phi}\sim 10^{13}$ GeV . 
From  Eq.(\ref{trh}) we find that 
$T_{RH} \simeq \mathcal{O}(1)$ GeV;
  a value much smaller than the values 
quoted in the standard GUT scenario ($T_{RH}\sim 10^{11}$). 
 This reheating temperature is of the same order or lower 
than the 
typical freeze-out temperature 
$T_F \simeq m_{\chi}/20\simeq O(1-5)$ GeV.

A detailed  analysis of the relic density with a low reheating
temperature has been considered in Ref. \cite{kolb} 
with different possible generic scenarios. 
The first scenario corresponds to a case where
the LSP is never in chemical equilibrium either before or after
reheating.
In the
 second the LSP's reach chemical
equilibrium but they freeze out before the completion of the
reheat process $T_F>T_{RH}$. Each of these scenarios leads to 
 different qualitative and quantitative predictions
for the relic density which can become quite different from the
standard computation we summarized in Eq.(\ref{omega1}).

CASE A) In case of non--equilibrium production and freeze out at the
early times and supposing that 
the number density of the LSP $n_{\chi}$ is much
smaller that $n_{\chi}^{eq}$, the Boltzmann equation
(\ref{boltzmann}) can be approximated and solved. 
In this case one gets \cite{kolb} 
\bea \Omega h^2 = 2.1 \times 10^4
\lambda_\star
\left( \alpha_s + \frac{\alpha_p}{4}
\right), \label{omega2} 
\eea 
where
\bea \lambda_\star = 
\frac{g^2}{4} \left(\frac{g_*(T_{RH})}
{10}\right)^{3/2} \left(\frac{10}{g_*(T_*)} \right)^3 \frac{(10^3
T_{RH})^7} {m_{\chi}^5},  
\eea 
and where $g$ is the number of degrees of
freedom of the LSP and $T_*$ is the temperature at which most of
the LSP production takes place, it is given by 
$T_* \sim 4 m_{\chi} /15$. 
As we can see this $\Omega h^2$ is proportional to
the annihilation cross section, instead of being inversely
proportional, as in Eq.(\ref{omega1}). However, the assumption
that $n_{\chi} << n_{\chi}^{eq}$ leads to a sever constraint on
the annihilation cross section~ \cite{kolb}: 
 $\alpha_s <
\bar{\alpha_s}$ and $\alpha_p < \bar{\alpha_p}$ where
${\alpha_s}\simeq 10^{-10}-10^{-9}$ and
${\alpha_p}\simeq 10^{-9}-10^{-8}$. 
So that for large
neutralino cross sections, which we are interested in,
eq.(\ref{omega2})  can not be applied. 

CASE B) With large annihilation
cross section 
 the LSP reaches equilibrium before reheating as
discussed in Ref.\cite{kolb} and its relic density $\Omega h^2$ is
very close to the one obtained by using the standard calculation
of relic density in Eq.(\ref{omega1}). 
In this case, 
$T_F$ is obtained by solving a different  equation as
before \cite{kolb}
\bea x_F = \ln \left[ C \left (
\alpha_s x_F^{5/2} + \frac{5}{4} \alpha_p x_F^{3/2} \right) \right ],
\eea 
where $C$ is an unimportant constant now.
The relic density
 $\Omega h^2$ is given by 
\be \Omega h^2 = 
\frac{g_*^{1/2}(T_{RH})}{g_*(T_F)} 
\frac{ 2.3 \times 10^{-11} T_{RH}^3 GeV^{-2}}
{m_{\chi}^3 (\alpha_s x_F^{-4} +4 \alpha_p x_F^{-5}/5)}.
\label{omega3} 
\ee
 In this case the relic density is again
inversely proportional to the annihilation cross section as in
Eq.(\ref{omega1}). Moreover, it has a further suppression due to
the very low reheating temperature effect. The typical value for
the relic density in this case for $\sigma^{ann} \sim 10^{-6}$
GeV$^{-2}$ is $\Omega h^2 \lsim 10^{-3}$. So it is even worse
result than the BB 
standard computation's one.

\section{Our Model}
In superstring
theory, which is the first motivation 
for  the idea of considering an  intermediate scale~\cite{interm1}, 
there are many moduli fields which acquire masses from 
SUSY breaking effects. 
Their masses are expected to
be of order the gravitino mass but their couplings with the MSSM
matter content are suppressed by a high energy scale~\cite{moroi}. 
The moduli ($\phi$) decay width will now be parameterized now 
as 
\be
\Gamma_{\phi} = \frac{1}{2 \pi} \frac{m_{\phi}^3}{M_I^2}
\label{gphi}
\ee
where $M_I$ is the unification scale ($\sim 10^{12}$ GeV) which acts
as an effective suppression scale. 

\begin{figure}[htb]
\centering
\epsfig{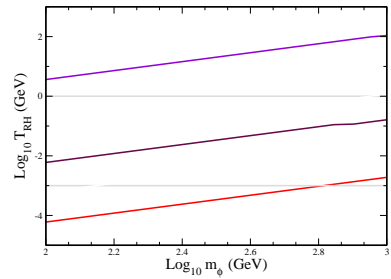}
\caption{ The reheating temperature $T_{RH}$ as a function of 
the  modulus mass $m_\phi$ for different values of the effective 
scale $M_I=10^{11},10^{14},10^{16}$ GeV. The horizontal 
at $O(1)$ MeV is the nucleosynthesis limit.}
\label{f1}
\end{figure}

The reheating temperature is obtained 
using eqs.~(\ref{trh}) and (\ref{gphi}) 
\bea 
\frac{T_{RH}}{ 5\ MeV} = 
\frac{ 10^{14}\ GeV}
{M_I}
\left(\frac{m_{\phi}}
{100\ GeV}\right)^{3/2}\
\left(\frac{10.75}{g_*(T_{RH})} \right)^{1/4}\ ,
\label{reheating}
\eea 
where, for example,
 $g_*=10.75$ for $T\sim$ $\cal{O}$($1-10$) MeV. 
This reheating temperature is 
shown as a function of the modulus mass in Fig.~\ref{f1} 
for different values of $M_I$.
The request that the modulus mass is larger than $\sim 100-500$ GeV in order 
to allow for kinematical 
decays into neutralinos of suitable mass $m_{\tilde\chi_1^0}\sim 50-200$ GeV,
limits in practice the reheating temperature 
to be above  $\sim 3$ GeV for the 
lowest scale on consideration $M_I=10^{11}$ GeV.
This has important consequences for the relic density computation.
As discussed in the introduction of this section,
we need 
a temperature smaller than the typical 
$T_F\simeq m_{\tilde\chi_1^0}/20\sim 3-10$ GeV, in
order to increase the relic neutralino density.
This implies that 
the scale $M_I\sim 10^{11}$ GeV is in the border of validity.
%Whereas values larger than this do work, e.g. 
On the other hand, for larger values we can obtain 
very easily interesting reheating temperatures. For example,
for $M_I=10^{12}$ GeV we have
$T_{RH}\gsim 0.3 $ GeV.
%, smaller values do not.
%we have to consider
%scales $M_I\gsim 10^{12}$ GeV, since
%$M_I= 10^{12}$ GeV corresponds to 
%$T_{RH}\gsim 0.3 $ GeV.
For the highest scale with an interesting
phenomenological value of the neutralino--nucleus cross section,
in the case of universality and moderate $\tan\beta$ \cite{interm1},
$M_I= 10^{14}$ GeV,
the lowest value of the reheating temperature corresponds to
$T_{RH}\sim 6 $ MeV. Larger values of the scale,
$M_I\gsim 10^{15}$ GeV, producing also a large
cross section, are possible in D-brane 
scenarios since non-universality in soft terms 
is generically present \cite{interm2,otrosneutrinos}.
In this case
the constraint $T_{RH} \gsim$ 1 MeV from 
nucleosynthesis can be translated into a constraint 
on the modulus mass $m_{\phi}\gsim 140$ GeV.

When considering the 
direct decay of the moduli fields into LSP (and therefore, an additional 
quantity of non 
thermal LSP production) the evolution of the cosmological 
LSP abundance becomes more complicated.
We have to consider now the coupled Boltzmann equations for the LSP, 
the moduli field and the radiation \cite{moroi,kolb}. The detailed
equations used by us appear explicitly in Ref.(\cite{torrenterelic}).
These equations depend on  
$m_\chi$, the LSP mass, $\bar{N}_{LSP}$,  the averaged number of LSP 
particles  produced in the decay of the modulus field.

Let us discuss qualitatively the solution following the arguments
used in ref.~\cite{moroi}.
For a $T_{RH}$ higher than $T_F$ the relic density will roughly
reproduce the usual result given by 
eq.~(\ref{thermal2}). However, for the interesting case for us
when $T_{RH}$ is lower than $T_F$, neutralinos produced
from modulus decay are never in chemical equilibrium, unlike the
thermal production case reviewed in Section~2.
As a consequence, its number density always decreases through
pair annihilation. When the annihilation rate 
$\langle\sigma^{ann}_{\tilde\chi_1^0} v \rangle n_{\tilde\chi_1^0}$
drops below the expansion
rate of the universe, $H$, 
the neutralino freezes out.
Then the relic density can be estimated as \cite{torrenterelic,moroi}
\begin{eqnarray}
\Omega_{\tilde\chi_1^0}
%^{(ann)} 
h^2 & =& \frac{3\ m_{\tilde\chi_1^0}\ \Gamma_{\phi}}
{2\ (2 \pi^2/45)\ g_\star\ T_{RH}^3\  \langle 
\sigma_{\tilde\chi_1^0}^{ann} v  \rangle}\ 
\frac{h^2}{ \rho_c /s_0}\ .
\label{nlsp}
\end{eqnarray}
This result is valid when there is a large number of neutralinos
produced by the modulus decay. When the number is insufficient,
they do not annihilate and therefore all the neutralinos survive.
The result in this case is given by 
\begin{eqnarray}
\Omega_{\tilde\chi_1^0}
%^{(0)} 
h^2 & =& \frac{3\ \bar N_{\tilde\chi_1^0}\ m_\chi\ \Gamma_{\phi}^2\ M_P^2}
{(2 \pi^2/45)\ g_\star\ T_{RH}^3\ m_{\phi}}\ 
\frac{h^2}{ \rho_c /s_0}\ .
\label{nlsp2}
\end{eqnarray}
%
%Basically, eq.~(\ref{nlsp}) is valid for $N_{\tilde\chi_1^0}\sim 1$,
%whereas eq.~(\ref{nlsp2}) is valid for 
%$N_{\tilde\chi_1^0}\lsim 10^{-3}-10^{-4}$ \cite{moroi}.
The actual relic density is estimated  \cite{moroi} as the minimum of 
(\ref{nlsp}) and (\ref{nlsp2}).

%%
%\begin{eqnarray}
%\Omega_{\chi} h^2 \sim min \left(\Omega_{\chi}^{(ann)} h^2,\ \Omega_{\chi}^{(0)} h^2\right)\ ,
%\end{eqnarray}
%%
Now we can apply the above equations to our case with
intermediate scales.
Using eqs.~(\ref{gphi}) and (\ref{reheating}), 
%and the current value of
%the critical density, $\rho_c/s_0=3.6\times 10^{-9}$ GeV$\times h^2$, 
we can write expressions (\ref{nlsp}) and (\ref{nlsp2})
as
\bea 
\Omega_{\tilde\chi_1^0} h^2 \propto  
\frac{ M_I m_{\tilde\chi_1^0}}
{\langle \sigma_{\tilde\chi_1^0}^{ann} v  \rangle m_\phi^{3/2} g_*^{1/4}}
\label{ann}
\eea 
\bea 
\Omega_{\tilde\chi_1^0} h^2 \propto 
\bar N_{\tilde\chi_1^0}
m_{\tilde\chi_1^0}
\frac{m_{\phi}^{1/2}}
{g_*^{1/4} M_I}
\label{(0)}
\eea 
From these equations 
%eq.~(\ref{ann}) 
we can see that even with a large annihilation cross section,
$\langle \sigma_{\tilde\chi_1^0}^{ann} v  \rangle\sim 10^{-8}$
GeV$^{-2}$, 
we are able to obtain the cosmologically interesting value  
$\Omega_{\tilde\chi_1^0}h^2\sim 1$.
For example for 
$M_I\sim 10^{13}$ GeV we obtain it 
%$\Omega_{\tilde\chi_1^0}\sim 0.1$
when $\bar N_{\tilde\chi_1^0}\sim 1$, using eq.~(\ref{ann}).

\begin{figure}[htb]
\centering
\begin{tabular}{c}
\psfig{file=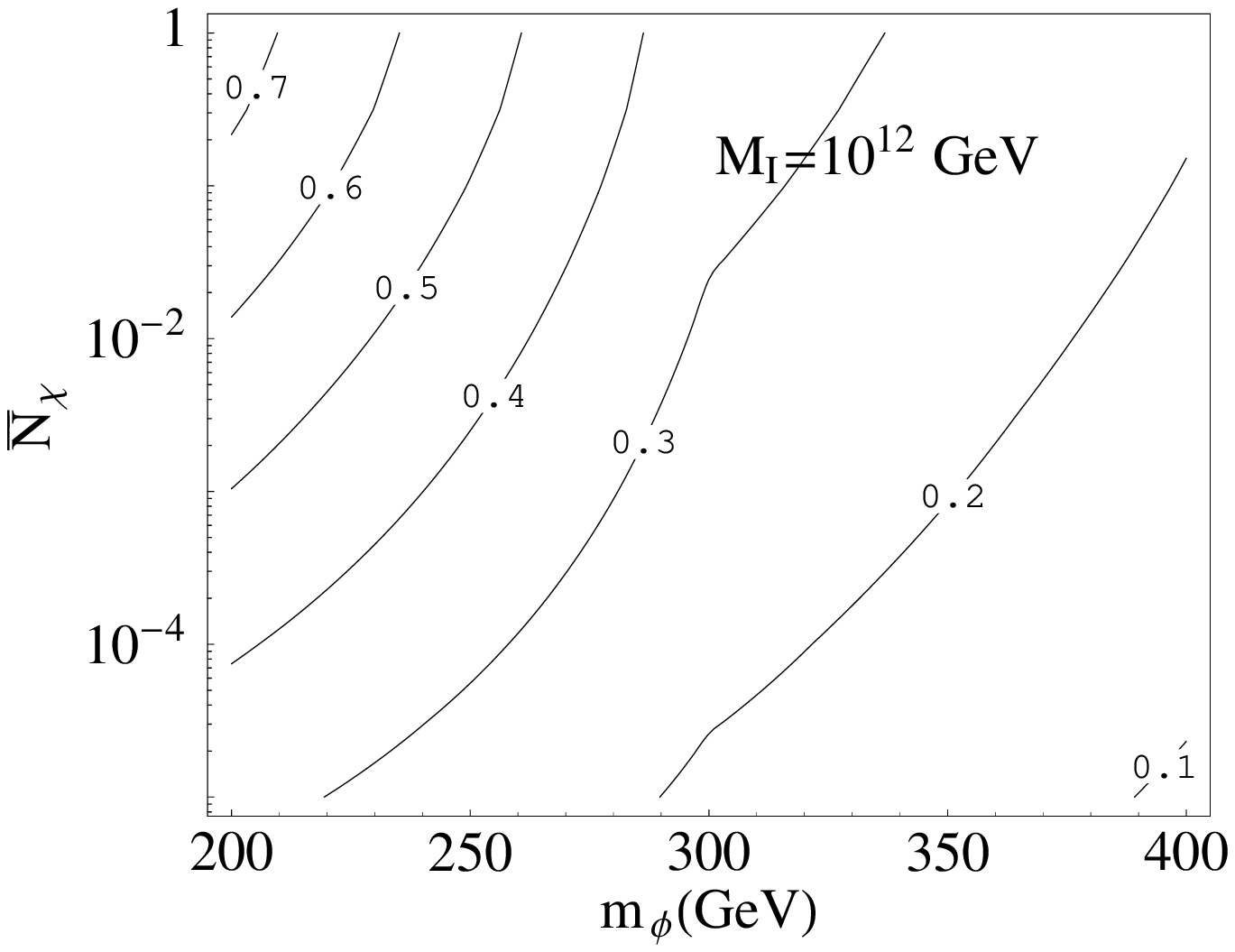,width=5cm} \\
\psfig{file=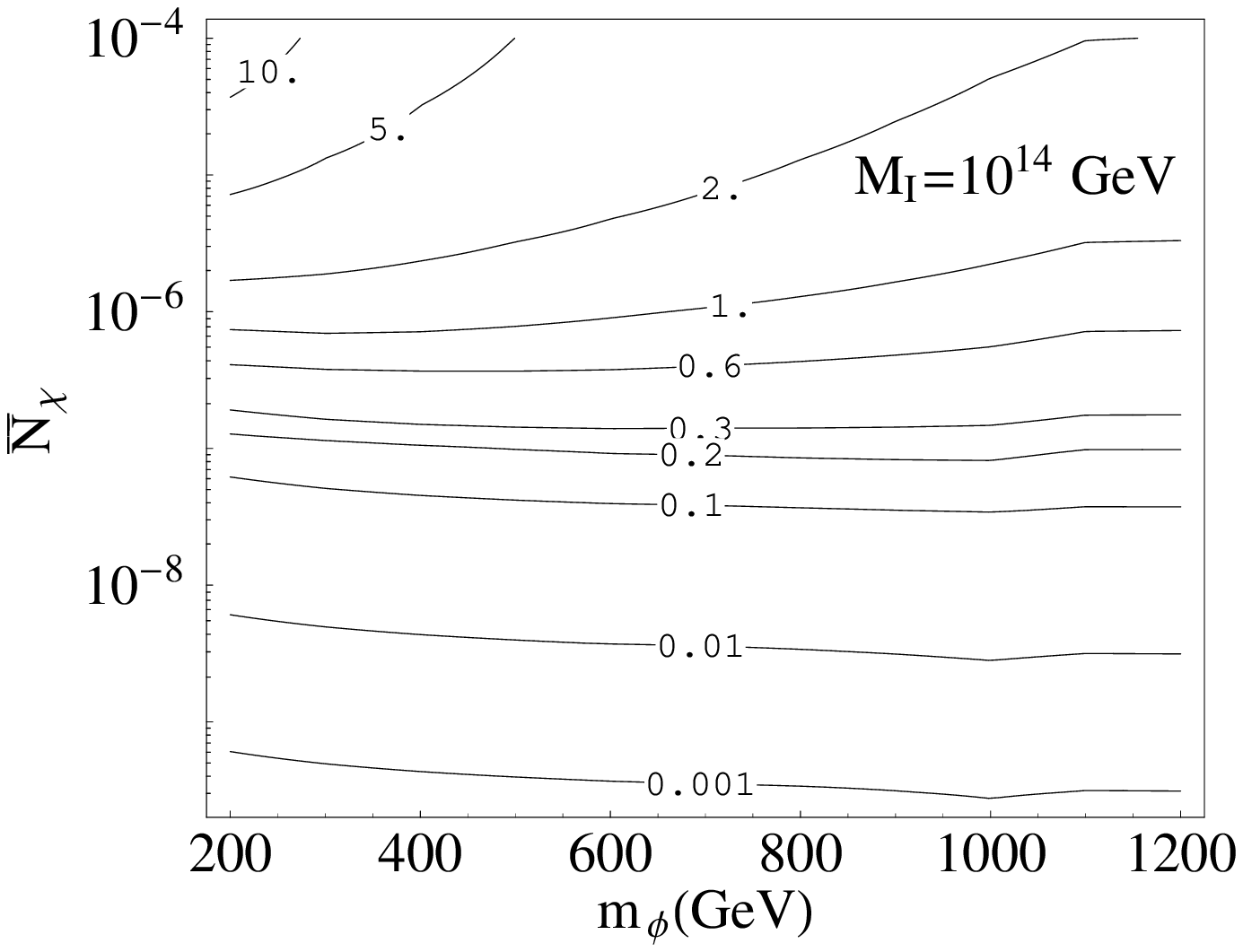,width=5cm} 
\end{tabular}
\caption{ 
Relic Neutralino density $\Omega h^2$ contours as a function of 
$m_\phi$ and $N$ for $m_\chi=100$ GeV and values of the 
intermediate scale $M_I=10^{11},10^{14}$ GeV.}
\label{f2}
\end{figure}

Previous expressions are useful but approximate. We have solved 
numerically the Boltzmann equations in several cases.
In Fig.~\ref{f2} we show in detail the results
of the numerical computation.
We have used a particular expression for the
  annihilation cross section introduced 
which is explained below.
In this figure, the contours of constant relic neutralino density
$\Omega_{\tilde\chi_1^0} h^2$
as a function of $m_\phi$
%$m_{\tilde\chi_1^0}$ 
and $\bar{N}_{\tilde\chi_1^0}$
are shown, for fixed values of $M_I$ and $m_{\tilde\chi_1^0}$.
In particular, we consider the cases
$M_I=10^{12}, 10^{13}, 10^{14}$ GeV, with $m_{\tilde\chi_1^0}=100$ GeV.
The
corresponding reheating temperatures can be obtained from Fig.~1.
 %to  
%$T_{RH}=??, ??, ??$ GeV,
%respectively.
Note that whereas many values of 
$\bar N_{\tilde\chi_1^0}$ correspond to a satisfactory relic density
for 
$M_I=10^{12}-10^{13}$ GeV,
for the case $M_I=10^{14}$ GeV only a small range works.

Let us finally finish this section with a few words about the 
neutralino annihilation cross section.
As it is well known,
in most of the parameter space 
of the MSSM the neutralino is mainly pure bino, and
as a consequence it will mainly annihilate into
lepton pairs through $t$--channel exchange  of right--handed
sleptons. The $p$--wave dominant cross section is given 
by \cite{olive,kolb}
\bea
\langle \sigma^{ann}_{\tilde\chi_1^0} v  \rangle 
\simeq 8\pi\alpha'^2
%\frac{g'^4}{2\pi}
\frac{1}{m_{\tilde\chi_1^0}^2}
\frac{1}{\left(1+x_{\tilde l_R}\right)^2}
\ \langle v^2  \rangle 
\ , 
\label{annihil} 
\eea 
where $x_{\tilde l_R}\equiv m_{\tilde l_R}^2/m_{\tilde\chi_1^0}^2$
and $\alpha'$ is the coupling constant for the $U(1)_Y$ interaction.
Taking $m_{\tilde l_R}\sim m_{\tilde\chi_1^0}\sim 100$ GeV,
$\langle \sigma^{ann}_{\tilde\chi_1^0} v  \rangle$ in 
eq.~(\ref{annihil}) becomes of the order of $10^{-9}$ GeV$^{-2}$ or smaller.
Using eq.~(\ref{thermal2}) an interesting relic abundance, 
$\Omega_{\tilde\chi_1^0}h^2\gsim 0.1$, 
is obtained.

However, in Refs.\cite{interm1,interm2}, we obtained that for 
$M_I\sim 10^{12}-10^{14}$ GeV the 
neutralino LSP acquires a sizeable Higgsino component, as we commented before this is 
essential for enhancing the LSP-nucleon cross section. 
For a Higgsino-like LSP the dominant annihilation channel
 is the decay into a W boson  pair. 
An upper bound on the cross section can be obtained \cite{moroi}:
\begin{eqnarray}
\langle \sigma_{ann} v\rangle & <=& \frac{\pi\alpha_2^2}{2}\frac{1}{m_\chi^2} \frac{(1-x_W)^{3/2}}
{(2-x_W)^2},
\end{eqnarray}
where $\alpha_2$ is the $SU(2)$ constant and 
$x_W=m_W^2/m_\chi^2$. 
It is this last expression for the cross section the one used in 
the numerical example presented in Fig.\ref{f1}.
For a LSP with 
$m_\chi\sim 100$ GeV, the formula gives a maximum $\sim 2\times 10^{-8}$ 
GeV$^{-2}$.
In the case that the LSP have sizeable components of both gaugino and Higgsino one expects 
smaller cross sections.

\section{Conclusions and Comments}

In conclusion we have analyzed
 the LSP relic density in supersymmetric models with
intermediate unification scale. 
we find that the reheating
temperature is of order 1 GeV ($M_I\sim 10^{12}$) and below down 
 to the nucleosynthesis threshold.
We have shown that
 in generic low $T_{RH}$ scenarios without non-thermal 
 direct production is expected that the LSP relic density is 
too low. As an alternative,
we have presented a concrete scenario where 
the reheating temperature is associated to 
the decay of oscillating moduli fields  appearing in string theories, 
we show that the LSP relic  density considerably increases with respect to 
the standard radiation-dominated case by the effect of the direct non-thermal 
production by the modulus field. The LSP can become a  good dark matter
candidate ($0.01-0.1 \gsim \Omega h^2 \lsim 0.3-1$) 
for $M_I\sim 10^{12}-10^{14}$ GeV and $m_\phi\sim 1-10$ TeV.

Let us finally remark that the numerical value of 
$\bar N_{\tilde\chi_1^0}$ is in general model dependent.
This was discussed in the context of supergravity in
ref.~\cite{moroi}. In this particular case
both values
$\bar N_{\tilde\chi_1^0}\sim 1$ and 
$\bar N_{\tilde\chi_1^0}\sim 10^{-3}-10^{-4}$ are plausible,
depending on the 
characteristics of the supergravity theory under consideration.

\end{document}